\documentclass[aps,pra,epsfig,twocolumn]{revtex4}
\usepackage{graphicx}
\usepackage{dcolumn} 
\usepackage{bm}
\begin{document}

\draft
\title{Engineering an interaction and entanglement between distant atoms}

\author{Stefano Mancini$^1$ and
Sougato Bose$^{2}$}

\affiliation{ $^1$ INFM \& Dipartimento di Fisica, Universit\`a di
Camerino, I-62032 Camerino, Italy
\\
$^2$Department of Physics and Astronomy, University College
London, Gower St., London WC1E 6BT, UK}

\date{\today}

\begin{abstract}
We propose a scheme to generate an effective interaction of
arbitrary strength between the internal degrees of freedom of two
atoms placed in distant cavities connected by an optical fiber.
The strength depends on the field intensity in the cavities. As an
application of this interaction, we calculate the amount of
entanglement it generates between the internal states of the
distant atoms. The scheme effectively converts entanglement
distribution networks to networks of interacting spins.
\end{abstract}

\pacs{Pacs No: 03.67.-a, 03.65.Ud, 32.80.Lg}

\maketitle


\section{Introduction}

It is known that two atoms separated by a large distance do not
interact directly with each other. Nonetheless, it would be highly
desirable to {\em engineer} a direct interaction between two such
atoms. To create such an interaction, one can try to artificially
set up a continuous exchange of real photons between the atoms in
a situation when virtual photons are not interchanged.  Here, we
propose such a scheme. We show how to generate an effective
interaction between atoms trapped in distant cavities connected by
optical fibers. This could be useful in generating entanglement
between the distant atoms. Entanglement shared between distant
sites is a valuable resource for quantum communications
\cite{bennett00}. Hence, we shall calculate the amount of
entanglement generated between the distant atoms by our engineered
interaction. Testing this entanglement will be equivalent to
testing the presence of a direct interaction between the distant
atoms. On a smaller scale, when the cavities are near, the scheme
would simply serve as an experiment to demonstrate the principle
that atoms trapped in distinct cavities can be made to directly
interact.

Numerous proposals have been made for entangling atoms trapped in
distinct cavities
\cite{zol3,Pel,zolb,sorensen98,ATOMS-meas,PAR,mancini01,duan,duan-kimble,simon,parkins,browne}.
Such atomic entanglement would be necessary to test Bell's
inequalities and quantum communication protocols with well
separated massive particles. In many cases an intermediate quantum
information carrier (such as a photon) between the atoms is
involved. Quantum information is mapped from atoms to photons in
one cavity and mapped back from photons to atoms in another. In
this paper, we eliminate the optical fields in the problem
altogether to obtain an effective {\em direct} interaction between
the distant atoms. This is, of course, much stronger than merely
generating entanglement. For example, a direct interaction, when
combined with local operations, can also be used to operate a
quantum gate directly between the distant atoms, to swap the
states of the atoms and so on. An Ising interaction, as generated
between the atoms in our case, can in fact, be used to construct a
universal quantum gate between the atoms (see the online
implementation associated with Ref.\cite{bremner}). Thus one can
use our method to directly link atomic qubits of distant quantum
processors.

\begin{figure}
\vspace*{-.5cm}
\includegraphics[width=1.5in, clip]{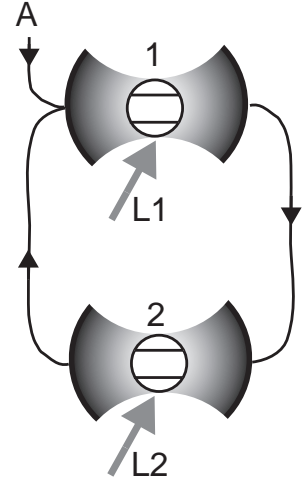}
\caption{Schematic description of the considered set-up. Two
distinct cavity $1$ and $2$ each containing a two-level atom are
connected via optical fibers. Practically the output of each
cavity enters the input of the other. Cavity $1$ also has an input
off-resonant driving field ${\cal A}$. $L1$ and $L2$ represent
resonant laser fields for local operations.} \label{fig1}
\end{figure}

\section{The Model}

We consider a very simple model consisting of two atoms, $1$ and
$2$, placed in distant cavities and interacting with light fields
in a dispersive way. The two cavities are then connected by
optical fibers as depicted in Fig.\ref{fig1}.

Interaction of atoms with light field in the
dispersive regime can be accounted for
by the following Hamiltonian
\cite{HOL91}
\begin{equation}\label{eq:Hint}
H_{int}=\chi A^{\dag}A\sigma_{1}^{(z)}
+\chi B^{\dag}B\sigma_{2}^{(z)}\,,
\end{equation}
where $A$, $B$ represent the relevant intracavity
radiation modes belonging cavity $1$ and $2$ respectively.
Furthermore,
$\sigma_{j}^{(x)}=(\sigma_{j}+\sigma_{j}^{\dag})$,
$\sigma_{j}^{(y)}=-i(\sigma_{j}-\sigma_{j}^{\dag})$
and $\sigma_{j}^{(z)}$ ($j=1,2$)
are the Pauli operators associated to the atomic internal degree
of freedom. The coupling constant $\chi$ (assumed, for the sake
of simplicity,  equal for the
two atoms) is given by
$g^{2}/\Delta$ with $g$ the dipole coupling and $\Delta$ the
detuning from the internal transition \cite{HOL91}.

Suppose, for the moment, that there is no connection between the
two cavities, and consider only the driving field of amplitude
${\cal A}$ at the first cavity. Then, the dynamics of the
intracavity modes is described by the
Langevin equations \cite{WM}
\begin{mathletters}
\begin{eqnarray}
\label{eqs:dyn}
 \dot A &=& -\left(\frac{\gamma}{2}+i\Delta\right)A
-i\chi A\sigma_1^{(z)}+\sqrt{\gamma}A_{in}+{\cal A}\,,
\\
\dot B &=& -\left(\frac{\gamma}{2}+i\Delta\right)B
-i\chi B\sigma_2^{(z)}+\sqrt{\gamma}B_{in}\,,
\end{eqnarray}
\end{mathletters}
where $A_{in}$, $B_{in}$ represent vacuum input noise and $\gamma$
is the cavity decay rate (assumed equal for the two cavities)
\cite{WM}. For the moment we have ignored the spontaneous decay
from the excited to the ground state of each atom. We will later
present a case with feasible parameters in which this is possible.

If we now connect the output of the cavity $1$ with the input
of cavity $2$, and the output of cavity $2$ with the input of
cavity $1$ (as in Fig.1), we will have additional dynamical terms
of the type \cite{WISE}
\begin{equation}\label{eqs:add}
\dot A = -\frac{\gamma}{2}A+\sqrt{\gamma}\,B'_{out}\,,
\quad
\dot B =
-\frac{\gamma}{2}B+\sqrt{\gamma}\,A'_{out}\,,
\end{equation}
where the subscript $out$ indicates the field outgoing a cavity,
while the prime sign means the retardation effect due to the
propagation of the field along the fiber, i.e., for a generic
operator ${\cal O}$ it is ${\cal O}'(t)\equiv{\cal O}(t-\tau)$
with $\tau$ the delay time. However, such effect can be described
as the introduction of a phase factor \cite{WISE}. Thus, adding
the terms (\ref{eqs:add}) to Eqs.(\ref{eqs:dyn}) and taking into
account the usual boundary conditions
$A_{out}\,(B_{out})=\sqrt{\gamma}A\,(B)-A_{in}\,(B_{in})$
\cite{WM}, we get
\begin{mathletters}
\begin{eqnarray}
\label{eqs:tot}
 \dot A &=& -\left(\gamma+i\Delta\right)A+\gamma
e^{i\phi_{12}}B -i\chi A\sigma_1^{(z)}
\nonumber\\
&&+\sqrt{\gamma}A_{in}-\sqrt{\gamma}
e^{i\phi_{12}}B_{in}+{\cal A}\,,
\\
\dot B &=& -\left(\gamma+i\Delta\right)B+\gamma e^{i\phi_{21}}A
-i\chi B\sigma_2^{(z)}
\nonumber\\
&&+\sqrt{\gamma}B_{in}-\sqrt{\gamma}
e^{i\phi_{21}}A_{in}\,,
\end{eqnarray}
\end{mathletters}
where $\phi_{12}$ ($\phi_{21}$) is the phase introduced along the
connection between the cavity $1$ ($2$) and the cavity $2$ ($1$).
Such phases can be experimentally controlled. For the moment we
ignore the loss effect in the fibers ({\em i.e.} assume lossless
fibers). Later we will analyze the case for a lossy fiber linking
the cavities.

Since we are interested on quantum effects at stationary regime,
we are going to linearize our equations. First, let us write the
steady state of the radiation fields by assuming  $\gamma$,
$\Delta > g$ ({\em i.e.} $\gamma > \chi$) and the expectation
values of the vacuum fields $A_{in},B_{in}$ to be much smaller
than the driving and cavity fields. It results
\begin{mathletters}\label{eqs:steady}
\begin{eqnarray}
\alpha&=&\frac{{\cal A}(\gamma+i\Delta)}{(\gamma+i\Delta)^{2}
-\gamma^{2}\exp\left[i\left(\phi_{12}+\phi_{21}\right)\right]}\,,
\\
\beta&=&\gamma\alpha\exp\left(i\phi_{21}\right)/
(\gamma+i\Delta)\,.
\end{eqnarray}
\end{mathletters}
Notice that the limit $\Delta\to 0$ and $\phi_{12}+\phi_{21}\to 0$
cannot be taken, since due to the recycling effect the intracavity
fields in such a case would explode.

Then, the linearized version of Eqs.(\ref{eqs:tot}) will be
\begin{mathletters}
\begin{eqnarray}
\label{eqs:totlin}
 \dot a &=& -\left(\gamma+i\Delta\right)a+\gamma
e^{i\phi_{12}}b -i\chi \alpha\sigma_1^{(z)}
\nonumber\\
&&+\sqrt{\gamma}a_{in}-\sqrt{\gamma}
e^{i\phi_{12}}b_{in}\,,
\\
\label{eqs:totlin1} \dot b &=&
-\left(\gamma+i\Delta\right)b+\gamma e^{i\phi_{21}}a -i\chi
\beta\sigma_2^{(z)}
\nonumber\\
&&+\sqrt{\gamma}b_{in}-\sqrt{\gamma}
e^{i\phi_{21}}a_{in}\,,
\end{eqnarray}
\end{mathletters}
where we have used the replacement $A\,(B)\to
\alpha\,(\beta)+a\,(b)$ and $a_{in}(b_{in})\equiv A_{in}(B_{in})$.
From Eqs.(\ref{eqs:totlin}) and (\ref{eqs:totlin1}), we can
adiabatically eliminate the radiation fields to obtain expressions
for $a$ and $b$ in terms of linear combinations of the Pauli
operators $\sigma_{1}^{(z)}$ and $\sigma_{2}^{(z)}$. In doing so
we can also neglect the noise terms for $ 1 << \gamma/\chi <<
|\alpha| $. Inserting the expressions for $a$ and $b$ (and hence
$A$ and $B$) in the Hamiltonian of Eq.(\ref{eq:Hint}), leads to an
effective interaction Hamiltonian for the two atoms of the type
\begin{equation}\label{eq:Heff}
H_{eff}=2J\sigma_{1}^{(z)}\sigma_{2}^{(z)}\,,
\end{equation}
with $J=\gamma\chi^{2}\Theta$, where we have assumed
\begin{eqnarray}\label{Theta}
\Theta&=&{\rm Im}\{\alpha^*\beta e^{i\phi_{12}}/
((\gamma+i\Delta)^{2}-\gamma^{2}e^{i(\phi_{12}+\phi_{21})}) \}\,,
\nonumber\\
&=& {\rm Im}\{\alpha\beta^{*}e^{i\phi_{21}}/
((\gamma+i\Delta)^{2}-\gamma^{2}e^{i(\phi_{12}+\phi_{21})})\}\,.
\end{eqnarray}
In deriving the above Hamiltonian (\ref{eq:Heff}) we have
neglected the self interaction terms since
$[\sigma_{j}^{(z)}]^2=1$. There are also additional local terms in
the Hamiltonian such as $\chi|\alpha|^2 \sigma_{1}^{(z)}$ and
$\chi|\beta|^2 \sigma_{2}^{(z)}$. Notice that the Hamiltonian
$H_{eff}$ is an {\em Ising Hamiltonian} whose spin-spin coupling
$J$ scales as radiation pressure and goes to zero for $\Delta\to
0$ and $\phi_{12}+\phi_{21}\to 0$.

We have thus managed to generate an effective Ising interaction
between two distant two-level atoms with the upper and lower
energy levels (say $|e\rangle_{j}$ and $|g\rangle_{j}$
with $j=1,2$)
taking the place of up and down spins of the
original Ising model. This interaction strength can be
arbitrarily increased by increasing the strength of radiation in
the cavities. This concludes the first part of our paper,
we next proceed to investigate an application
of this interaction to entangling the distant atoms.

\section{Entanglement}
Gunlycke {\em et al.} have recently investigated thermal
entanglement in the Ising model in an arbitrarily directed
magnetic field \cite{gunlycke01}. In particular, it was shown in
Ref.\cite{gunlycke01} that to get entanglement in the Ising model,
it is necessary to have a magnetic field perpendicular to the $z$
direction. To this end, we apply local laser fields to each atom
(L1 and L2 of Fig.\ref{fig1}) such that the local Hamiltonian
$H_{local}$ given by
\begin{equation}\label{eq:Hlocal}
H_{local}=B \sigma_1^{(x)} + B \sigma_2^{(x)},
\end{equation}
acts on the atoms in addition to $H_{eff}$. It is also assumed
that the local terms of the effective Hamiltonian ($\chi|\alpha|^2
\sigma_{1}^{(z)}$ and $\chi|\beta|^2 \sigma_{2}^{(z)}$) are fully
cancelled by choosing an appropriate detuning of the local laser
fields from the $|e\rangle_{j}\rightarrow|g\rangle_{j}$
transition. We choose $B=\eta J$, with $\eta \ll 1$ so that the
earlier derivation of the effective Ising Hamiltonian is
unaffected by the presence of these extra classical laser fields.
Thus the total Hamiltonian of the system is
\begin{equation}\label{eq:Htot}
H_{tot} = H_{local} + H_{eff}.
\end{equation}

The Hamiltonian $H_{tot}$ has the following eigenvectors:
\begin{mathletters}
\begin{eqnarray}
|\psi_{1}\rangle &=&
\frac{\eta}{2\sqrt{1+\eta^{2}+\sqrt{1+\eta^{2}}}}
\left(|g\rangle_{1}|g\rangle_{2}+|e\rangle_{1}|e\rangle_{2}\right)
\nonumber\\
&& - \frac{1+\sqrt{1+\eta^{2}}}
{2\sqrt{1+\eta^{2}+\sqrt{1+\eta^{2}}}}
\left(|e\rangle_{1}|g\rangle_{2}+|g\rangle_{1}|e\rangle_{2}\right)\,,
\nonumber\\
\\
|\psi_{2}\rangle&=&\frac{1}{\sqrt{2}}\left(
|e\rangle_{1}|g\rangle_{2}-|g\rangle_{1}|e\rangle_{2}\right)\,,
\\
|\psi_{3}\rangle&=&\frac{1}{\sqrt{2}}\left(
|g\rangle_{1}|g\rangle_{2}-|e\rangle_{1}|e\rangle_{2}\right)\,,
\nonumber\\
|\psi_{4}\rangle &=&
\frac{\eta}{2\sqrt{1+\eta^{2}-\sqrt{1+\eta^{2}}}}
\left(|g\rangle_{1}|g\rangle_{2}+|e\rangle_{1}|e\rangle_{2}\right)
\nonumber\\
&& - \frac{1-\sqrt{1+\eta^{2}}}
{2\sqrt{1+\eta^{2}-\sqrt{1+\eta^{2}}}}
\left(|e\rangle_{1}|g\rangle_{2}+|g\rangle_{1}|e\rangle_{2}\right)\,,
\nonumber\\
\end{eqnarray}
\end{mathletters}
with eigenvalues
$E_{1}=-2\sqrt{B^{2}+J^{2}}$,
$E_{2}=-2 J$, $E_{3}=2 J$ and
$E_{4}=2\sqrt{B^{2}+J^{2}}$.

Let us now consider as initial state of the two atoms
the ground state $|g\rangle_{1}|g\rangle_{2}$, then
we can expand it
over the eigenstates basis as
\begin{equation}\label{Psi0}
|\Psi(0)\rangle=|g\rangle_{1}|g\rangle_{2}=\sum_{j=1}^{4}
C_{j}|\psi_{j}\rangle\,,
\end{equation}
with
\begin{mathletters}
\begin{eqnarray}
C_{1}&=&-\frac{\left(1-\sqrt{1+\eta^{2}}\right)
\sqrt{1+\eta^{2}+\sqrt{1+\eta^{2}}}}
{2\eta\sqrt{1+\eta^{2}}}\,,
\\
C_{2}&=&0\,,
\\
C_{3}&=&\frac{1}{\sqrt{2}}\,,
\\
C_{4}&=&\frac{\left(1+\sqrt{1+\eta^{2}}\right)
\sqrt{1+\eta^{2}-\sqrt{1+\eta^{2}}}}
{2\eta\sqrt{1+\eta^{2}}}\,.
\end{eqnarray}
\end{mathletters}
The evolution of the state (\ref{Psi0})
under $H_{tot}$ will give
\begin{eqnarray}\label{Psit}
|\Psi(t)\rangle&=&C_{1}e^{2i\tau\sqrt{1+\eta^{2}}}|\psi_{1}\rangle
\nonumber\\
&&+C_{3}e^{-2i\tau}|\psi_{2}\rangle
\nonumber\\
&&+C_{4}e^{-2i\tau\sqrt{1+\eta^{2}}}|\psi_{4}\rangle\,,
\end{eqnarray}
where we have introduced the scaled time $\tau=J t$.

\begin{figure}
\begin{center}
\includegraphics[width=3.5in, clip]{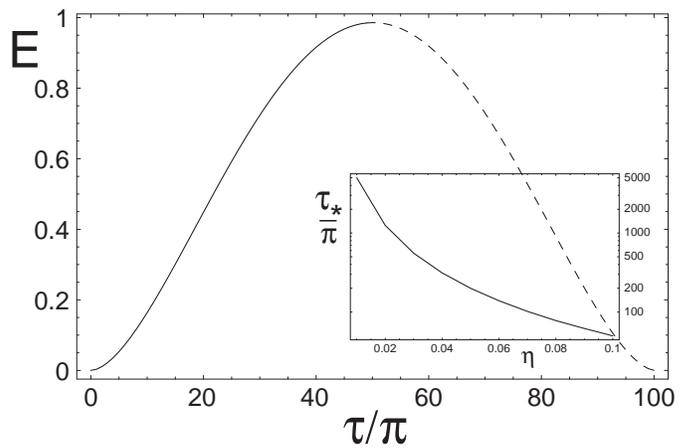}
\caption{The figure shows the plot of the amount $E$ of
entanglement between the distant atoms versus the scaled time
$\tau$ for $\eta=0.1$. The inset shows, on a $\log$ scale, the
time $\tau_{*}$ versus $\eta$. } \label{fig2}
\end{center}
\end{figure}

In Fig.\ref{fig2} we have plotted the entanglement of formation
$E$ (from the formula by Wootters \cite{wootters98}) for the state
(\ref{Psit}) as a function of $\tau$. We note that $E$ can
approach the maximum value $1$ before diminishes. Its behavior is
quasiperiodic as we are considering Hamiltonian dynamics with
incommensurable frequencies. As soon as $E$ reaches a value
$\simeq 1$ we can suppose to turn off $H_{local}$ (or $H_{tot}$)
and leave the atoms in a maximally entangled states. Notice, from
Eq.(\ref{Psit}), that entanglement does not directly depend on the
interaction strength, but rather on the ratio $B/J$. As a
consequence $\eta$ determines the value of time for which the
maximal entanglement ($E_{\mbox{max}}$) is reached. Setting
$\tau_{*}$ as the smaller value of $\tau$ such that $E\approx
E_{\mbox{max}}(\eta)\approx 1$, the inset of Fig.\ref{fig2} shows
how $\tau_{*}$ increases by diminishing the value of $\eta$.

\section{Discussion}
We have completely {\em eliminated} the optical field in
the process of deriving the effective Hamiltonian. In doing that
we have also neglected the losses along the fibers.
We now examine what happens if the fiber is lossy. The important
effect of a lossy fiber is that the primed fields $A^{'}_{out}$
and $B^{'}_{out}$ also have a damping term (say
$\exp{(-\Gamma_f)}$) in addition to the phase factor
$\exp{(i\phi_{12})}$ relative to their unprimed counterparts.
Hence we should be able to model the effect of a lossy fiber
phenomenologically by replacing $\exp{(i\phi_{12})}$ and
$\exp{(i\phi_{21})}$ by $\exp{(i\phi_{12}-\Gamma_f}$ and
$\exp{(i\phi_{21}-\Gamma_f)}$ respectively in Eqs.(\ref{eqs:tot})
to (\ref{eq:Heff}). When this replacement is done, only the
dependence of $a$ on $\sigma_2^{(z)}$ and of $b$ on
$\sigma_1^{(z)}$ is affected ($a$ and $b$ still depend in the same
way on the local terms). The dependence of $J$ on $\Gamma_f$ is
the found, in general, to be quite complicated (depends on the
explicit values of $\gamma,\Delta,\phi_{12}$ and $\phi_{21}$).
However, if we make the simplifying assumption that
$\Delta>>\gamma$, then $J$ is simply replaced by
$J\exp{(-2\Gamma_f)}$. In the typical optical fibers used today,
the loss rates are as low as $0.35$ dB per kilometer (this data is
from a quantum communication experiment with photons
\cite{gisin}). This translates to $\Gamma_f\approx 0.08$ for a
fiber of one kilometer (separating cavities by the same distance).
Then the coupling strength $J$ between the atoms is about $92$
percent of that estimated by Eq.(\ref{eq:Heff}).

  Finally, once generated, entanglement may have
a stability problem due to the atomic decay from the excited
states. However, one can deal this problem by using, as
$|e\rangle$ and $|g\rangle$, Zeeman ground state levels in a
$\Lambda$ configuration \cite{Parkins93}. This guarantees long
lived states and its use has been already proposed within quantum
computation \cite{Pel95}. A recent experiment with atoms in
optical cavities has used precisely this type of atomic system
\cite{kuhn} and we will estimate the feasibility of our proposal
by slightly modifying the parameters of that experiment. The
strength $\chi$ of our scheme is given in terms of two single
photon Rabi frequencies $g$ and $\Omega$ and an atomic detuning
$\Delta_{a}$ (different from our cavity detuning $\Delta$) of
Ref.\cite{kuhn} as $\chi=g\Omega/\Delta_{a}$. The parameters of
the experiment of Ref.\cite{kuhn} are
$(g,\Omega,\Delta_{a},\gamma)=2\pi(2.5,8,-20,1.25)$ MHz. We
increase $\gamma$ five times, which is easy to do (higher cavity
decay rate) and choose $\Delta<<\gamma$ and
$\phi_{12}=\phi_{21}=\pi/4$ for simplifying the expression of $J$
(these are not necessary). Then we have $J\approx \chi^2
\bar{n}/2\gamma\approx \bar{n}/2$~MHz (where $\bar{n}=|\alpha|^2$
is the number of photons in the first cavity). Thus with
$\bar{n}\sim 50-100$, we already have $J\sim 25-50$MHz, which is
indeed a strength of interaction comparable to usual atom-light
interaction strength in cavities.

\section{Conclusions}
In conclusion, we have presented a scheme for generating an
Ising interaction between distant atoms. Such a scheme has greater
potential than any scheme that merely entangles the distant atoms.
For example, a direct interaction can be used to implement a two
qubit logic gate between the distant atoms. The strength of the
coupling can be made arbitrary by pumping more or less radiation
into any of the cavities. This is a result of using off-resonant
coupling, between each atom and its cavity mode. The coupling of
light with any general macroscopic object, called ``ponderomotive"
coupling (see Refs.\cite{mancini01} for its applications in the
context of entanglement) is of the {\em same} type. Thus our
entangling scheme could potentially be extended to generate
thermal entanglement for macroscopic objects \cite{pond}.


\end{document}